\documentstyle[12pt,epsfig]{article}
\newcommand{\scaption}[1]{\caption{\protect{\footnotesize  #1}}}

\newcommand{\av}[1]{\mbox{$ \langle #1 \rangle $}}

\newcommand{\W}{\mbox{$W~$}}

\newcommand{\xB}{\mbox{$x~$}}  
\newcommand{\xb}{\mbox{$x~$}}  

\newcommand{\Qsq}{\mbox{$Q^2~$}}
\newcommand{\et}{\mbox{$E_T~$}}
\newcommand{\kt}{\mbox{$k_T~$}}
\newcommand{\pt}{\mbox{$p_T~$}}
\newcommand{\ptmax}{\mbox{$p_T^{\rm max}~$}}

\newcommand{\GeV}{\mbox{\rm ~GeV~}}
\newcommand{\GeVx}{\rm GeV}

\newcommand{\GeVsq}{\mbox{${\rm ~GeV}^2~$}}

\newcommand{\ep}{\mbox{$ep~$}}

\begin{document}
\begin{titlepage}

%
\noindent
{\tt DESY 96-215    \hfill    ISSN 0418-9833} \\
{\tt October 1996}                  \\

\begin{center}

\vspace*{2cm}

\begin{Large}

{\bf  Measurement of Charged Particle Transverse Momentum Spectra
      in Deep Inelastic Scattering
   }\\[1.5cm]

\vspace*{2.cm}
H1 Collaboration \\
\end{Large}

\vspace*{1cm}

\end{center}

\vspace*{1cm}

\begin{abstract}

\noindent
Transverse momentum spectra of charged particles produced in 
deep inelastic scattering are measured
as a function of the kinematic variables
\xB and \Qsq using the H1 detector at the \ep collider HERA.
The data are compared to different parton emission models,
either with or without ordering of the emissions 
in transverse momentum.
The data provide evidence for a relatively large 
amount of parton radiation between the current and the remnant systems.

\vspace{1cm}

\end{abstract}
\end{titlepage}

\begin{Large} \begin{center} H1 Collaboration \end{center} \end{Large}
\begin{flushleft}
 C.~Adloff$^{35}$,                
 S.~Aid$^{13}$,                   
 M.~Anderson$^{23}$,              
 V.~Andreev$^{26}$,               
 B.~Andrieu$^{29}$,               
 A.~Babaev$^{25}$,                
 J.~B\"ahr$^{36}$,                
 J.~B\'an$^{18}$,                 
 Y.~Ban$^{28}$,                   
 P.~Baranov$^{26}$
% ...

\end{flushleft}
\begin{flushleft} {\it
% ...
$ ^{13}$ II. Institut f\"ur Experimentalphysik, Universit\"at Hamburg,
                                     Hamburg, Germany$^ a$  \\
% ...
$ ^{18}$ Institute of Experimental Physics, Slovak Academy of
                Sciences, Ko\v{s}ice, Slovak Republic$^{f, j}$\\
% ...
$ ^{23}$ Physics Department, University of Manchester,
                                          Manchester, UK$^ b$\\
% ...
$ ^{25}$ Institute for Theoretical and Experimental Physics,
                                                 Moscow, Russia \\
 $ ^{26}$ Lebedev Physical Institute, Moscow, Russia$^ f$ \\
% ...
$ ^{28}$ LAL, Universit\'{e} de Paris-Sud, IN2P3-CNRS,
                            Orsay, France\\
 $ ^{29}$ LPNHE, Ecole Polytechnique, IN2P3-CNRS,
                             Palaiseau, France \\
% ...
$ ^{35}$ Fachbereich Physik, Bergische Universit\"at Gesamthochschule
               Wuppertal, Wuppertal, Germany$^ a$ \\
 $ ^{36}$ DESY, Institut f\"ur Hochenergiephysik,
                              Zeuthen, Germany$^ a$\\
% ...
$ ^a$ Supported by the Bundesministerium f\"ur Bildung, Wissenschaft,
        Forschung und Technologie, FRG,
        under contract numbers 6AC17P, 6AC47P, 6DO57I, 6HH17P, 6HH27I,
        6HD17I, 6HD27I, 6KI17P, 6MP17I, and 6WT87P \\
 $ ^b$ Supported by the UK Particle Physics and Astronomy Research
       Council, and formerly by the UK Science and Engineering Research
       Council \\
% ...
$ ^f$ Supported by the Deutsche Forschungsgemeinschaft \\
% ...
$ ^j$ Supported by VEGA SR grant no. 2/1325/96 \\

   } \end{flushleft}          
\newpage

\section{Introduction}

\begin{figure}[htp]
   \centering
   \vspace{-0.2cm}
   \epsfig{file=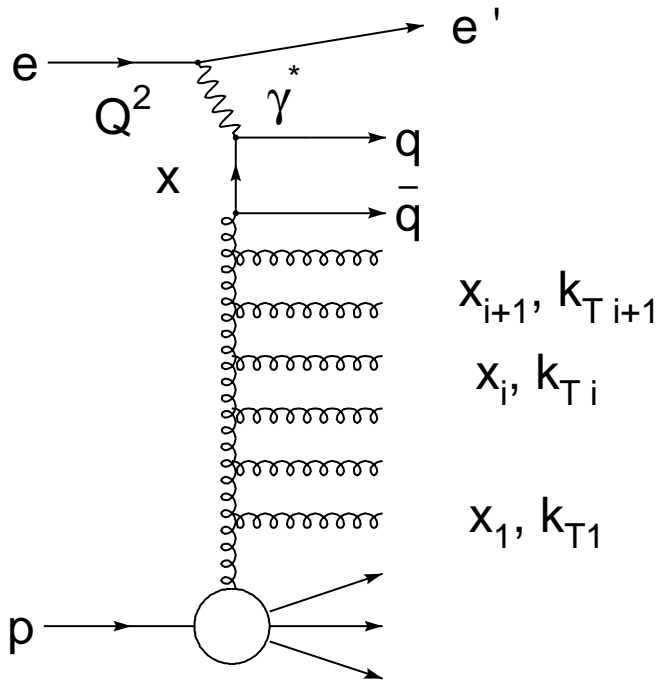,width=10cm}
   \caption{\em 
      Generic diagram for parton evolution. }
   \label{cascade} 
\end{figure}

The electron-proton collider HERA allows to explore a new kinematic
region in deep inelastic scattering (DIS) down to the very small
Bjorken-\xb of about  $10^{-5}$, where new dynamic features of 
QCD may show up.
So far,
 the description of the nucleon structure function 
measurement 
by perturbative QCD, cast into the
DGLAP (Dokshitzer-Gribov-Lipatov-Altarelli-Parisi) parton evolution 
equations \cite{dglap}, has been extremely successful, and constitutes
one of the major successes of QCD. 
At small enough $x$, however,
these equations are expected to 
cease to be a good approximation.
For the small \xb regime the 
BFKL 
(Balitsky-Fadin-Kuraev-Lipatov) equation \cite{bfkl} has been
suggested, and  
it would be very interesting to test QCD in such a new regime. 
To lowest order the BFKL and DGLAP equations resum the leading 
logarithmic $(\alpha_s \ln (1/x))^n$ 
and $(\alpha_s \ln (Q^2/ Q_0^2))^n$ contributions
respectively, where $Q^2$ is the invariant mass squared of the virtual photon
and $Q_0$ is the cut-off for the perturbative evolution.
 In these approximations the leading diagrams are of the
ladder type (Fig.~\ref{cascade}). 
The leading log
DGLAP resummation corresponds to a strong
ordering of the transverse momenta \kt (w.r.t. the proton beam)
in the parton cascade
($Q_0^2 \ll k_{T1}^2 \ll ... k_{Ti}^2 \ll ... Q^2$).
In the BFKL regime, the transverse momenta follow
a kind of random walk
($k_{Ti}^2 \approx k_{Ti+1}^2$)~\cite{ordering}. 

The strong rise towards small \xb
of the structure function $F_2$ measured at HERA \cite{h1f2,zeusf2}
can be described with DGLAP evolution \cite{akms,grv} and is also
consistent with the BFKL expectation.
Less inclusive measurements, namely 
of the hadronic final state emerging from the cascade,
may offer  
more sensitive means of searching for BFKL evolution.
One possible signature is the quantity of 
transverse energy, $E_T$, produced in a 
region between the current (struck) quark and the 
proton remnant \cite{h1flow2}. 
As a consequence of the strong \kt ordering, DGLAP
evolution is expected to produce less \et 
than BFKL evolution \cite{durham}.
The HERA \et flow data \cite{h1flow3} can 
be interpreted consistently within the BFKL framework,
but from these data alone it 
is not possible to disentangle non-perturbative hadronization effects
from perturbative parton radiation, precluding 
an unambiguous test of the underlying parton dynamics 
\cite{kuhlen1,sci}. In fact, the data can also be described
with a DGLAP based parton shower model with the assumption 
that the 
hadronization effects are large~\cite{sci}.

In this paper complementary measurements of charged particle 
transverse momentum ($p_T$) spectra are presented.
Such spectra have
been proposed~\cite{kuhlen2}
as a more direct probe of the underlying 
parton dynamics than the \et flow measurements.
It has been shown with the aid of QCD models that the high-$p_T$ tail is sensitive to parton 
radiation, whereas the production of hard particles from hadronization
is suppressed. For the $k_T$ unordered scenario a harder $p_T$ 
spectrum is expected in the central region than for the $k_T$ 
ordered evolution, because parton emissions are less 
restricted in $k_T$. The unsuppressed gluon radiation should give
rise to a hard tail of the $p_T$ distribution for the hadrons 
emerging from the cascade. The tail would be absent if the $E_T$ seen
in the data stems predominantly from hadronization. 
A particular good discrimination can be obtained with events in which
$E_T$ is large. Radiation of a hard gluon may lead to large $E_T$ and 
a high $p_T$ particle, but if hadronization were responsible for 
the large $E_T$, many soft particles are expected instead.
Furthermore, a different $x$-behaviour of the $p_T$ spectra is 
predicted. The phase space increases with $W^2 \approx Q^2/x$ 
($W$ is the invariant mass of the hadronic final state) and
therefore one may expect more gluon radiation, more $E_T$ and harder
$p_T$ spectra with decreasing $x$, but not if $k_T$ ordering or another
mechanism is invoked to suppress gluon radiation.
The measurements presented here are compared to 
predictions from Monte Carlo models with
different mechanisms for gluon radiation.
They incorporate QCD evolution in
different approximations and utilize phenomenological models
for the non-perturbative hadronization phase.

The MEPS model (Matrix Element plus Parton Shower) 
incorporated within the LEPTO generator~\cite{lepto}, 
uses the first order QCD matrix elements, 
with additional
soft emissions generated by adding leading log DGLAP parton showers.
Though this approach 
is widely used with considerable success,
it has to be pointed out that the DGLAP formalism was derived
for the totally inclusive structure function evolution,
and it can be questioned whether it is fully applicable
for exclusive final states \cite{sci}.
In the colour dipole model (CDM)~\cite{dipole}, as implemented
in ARIADNE \cite{ariadne},
radiation stems from
colour dipoles formed by 
the colour charges. 
Both programs
use the Lund string model \cite{string} for hadronizing the 
partonic final state.
The HERWIG model \cite{herwig}
is also based on leading log parton showers, with
additional matrix element corrections \cite{seymour}.
This model implements an alternative 
(cluster) fragmentation scheme \cite{cluster}.
The CDM description 
of gluon emission is similar to that of BFKL evolution 
to the extent that 
the gluons emitted by the dipoles
do not obey strong ordering in $k_T$~\cite{bfklcdm}. 
In the MEPS and HERWIG models 
the partons emitted in the cascade
are strongly ordered in $k_T$, because they are based on 
leading log DGLAP parton showers. 
The models are used in their versions LEPTO 6.4 for MEPS,
ARIADNE 4.08 for CDM and HERWIG 5.8, together 
with the parton density parameterization 
GRV \cite{grv}.
They
provide a satisfactory
overall description of current DIS final state data \cite{carli},
in particular of the \et flows~\footnote{
However, in LEPTO (among other things) the new concept of 
soft colour interactions \cite{sci} 
had to be introduced to reproduce the level of \et seen in the data
\cite{h1flow2,carli,sci}. Intriguingly,
this mechanism also produces 
rapidity gap events \cite{gap} at a rate of about 10\%,
comparable to observation~\cite{sci}.
Also the cluster fragmentation scheme in HERWIG
produces rapidity gap events.
In this paper rapidity gap events are excluded.}.
For this analysis the use of the
alternative parameterization MRS-H \cite{mrsh} 
would give model predictions which differ only slightly from 
those shown. 
       
\section{Detector Description}

A detailed description of the H1 apparatus  
can be found elsewhere~\cite{h1nim}.
The following briefly describes the detector components relevant 
to this 
analysis. 
 
The hadronic energy flow and the scattered positron are measured with a
liquid argon~(LAr) calorimeter and a  
backward electromagnetic lead-scintillator calorimeter (BEMC) 
respectively.
The LAr calorimeter~\cite{larc}  extends over the polar angular range
$4^\circ < \theta <  153^\circ$ with full azimuthal coverage, where
 $\theta$ is defined with respect to the proton
beam direction ($+z$ axis).
The depth 
varies between 4.5 and 8 interaction lengths for $\theta <  125^\circ$.
Test beam measurements of the LAr~calorimeter modules show an
energy resolution 
of $\sigma_{E}/E\approx 0.50/\sqrt{E\;[\GeVx]} \oplus 0.02$  for 
charged pions~\mbox{\cite{h1pi}}.
The 
absolute scale of the hadronic energy measurement
is presently known to $5\%$, as
determined from studies of the 
transverse momentum ($p_T$) balance in DIS events.

The BEMC \cite{h1bemc} (with a depth of 22.5 radiation lengths and 1
interaction length) covers the backward region,
 $151^\circ < \theta < 177^\circ$.
A major task of the BEMC is to trigger and to measure precisely
 scattered positrons
in DIS processes with $Q^2$ values ranging from 5 to 100 GeV$^{2}$. 
The BEMC energy scale for electrons is known to an accuracy of $1\%$.
Its resolution is given by 
$\sigma_{E}/E = 0.10/\sqrt{E\;[\GeVx]} \oplus 0.42/E[\GeVx] \oplus 0.03$
\cite{h1bemc}.  
A backward proportional chamber (BPC), in front of the BEMC, with an angular
acceptance of $155.5^\circ < \theta < 174.5^\circ$, serves to identify electrons
and to measure precisely their direction. 
Using information from the BPC, the BEMC and the reconstructed event vertex, the
polar angle of the scattered positron is known to a precision of 1 mrad. 
 
The calorimeters are 
surrounded by a superconducting solenoid which provides a uniform
magnetic field of $1.15$ T parallel to the beam axis in the tracking region.

The tracking system consists of a central and a forward part.  
The central tracking system is mounted concentrically around the 
beam line, covering polar angles of $20^\circ < \theta  < 160^\circ$.
Charges and momenta of charged particles are measured
by two coaxial cylindrical drift chambers 
(central jet chamber, CJC), providing up to 56 space points in 
the radial plane. Longitudinal coordinates are obtained via charge
division from the CJC, 
and from dedicated drift chambers which are interleaved
with the CJC.
For the CJC the resolutions achieved are 
$\sigma_{p_T}/p_T \approx 0.009\cdot p_T\;[\GeVx] \oplus 0.015$
and $\sigma_\theta = 20$~mrad \cite{h1nim}.
The central tracking system is complemented by the forward tracker,
which is built in three sections covering a polar
angle range of $7^\circ < \theta  < 25^\circ$.  
Each section is made up of a series of 
multiwire proportional and drift chambers which are arranged to 
facilitate track reconstruction  
at small angles to the proton beam.  
For tracks fitted to a vertex 
the resolution has been shown to be 
$\sigma_{p_T}/p_T \approx 0.02\cdot p_T\;[\GeVx]/\sin\theta \oplus 0.1$
and the angular resolution, $\sigma_\theta$, 
to be better than 1 mrad \cite{ftracker}.

Scintillation counters installed behind the BEMC are used to reject 
proton induced background. 
Small angle electron/photon taggers are used for luminosity measurements
and for the study of photoproduction background.

\section{
Event and Track Selection}

The data used in this analysis were collected in 1994, 
with positrons of energy $E_e=27.5\GeV$ colliding with
protons of energy $E_p=820$~GeV, resulting in a total centre
of mass energy of $\sqrt{s}=300 \GeV$.  
The data correspond to an integrated luminosity of 1.3 pb$^{-1}$.
For this analysis DIS events with 
$5<Q^2 < 50$~GeV$^2$ are used, in which the scattered positron is observed in
the BEMC.  
The events are triggered by requiring a cluster of more than 4~GeV
in the BEMC.
After reconstruction, DIS events are selected in
the following way: 

\begin{itemize}
\item The scattered positron, defined as the most energetic BEMC cluster,
  must have an energy $E'_e$ larger than 12~GeV and a polar angle $\theta_e$
  below $173^\circ$ in order to ensure high trigger efficiency and a small
  photoproduction background~\cite{h1f2}. 

\item The lateral size of the positron
  cluster, calculated as the energy weighted radial distance of the cells from
  the cluster centre, has to be smaller than 4~cm.
  The cluster must be associated
  with a reconstructed BPC space point 
  Further reduction of photoproduction background 
  and the removal of events in which an energetic photon is radiated off 
  the incoming positron 
  is achieved by 
  requiring $\sum_j{(E_j-p_{z,j})} >30\GeV$ \cite{h1f2},
  with
  the sum extending over all particles $j$ (measured calorimetrically)
  in the event.

\item 
  The radial coordinate of the BPC hit must be less than
  60~cm, corresponding
  to a positron angle above $157^\circ$ with respect to the nominal
  interaction point, ensuring full containment of the positron shower
  in the BEMC. 

\item 
  The $z$ position of the  
  event vertex reconstructed from charged tracks has to be within 30~cm
  of the average of all collision events. 

\item The energy in the forward region  
  ($4.4^\circ < \theta < 15^\circ$) has to be larger than 0.5~GeV in order
  to exclude diffractive events with large rapidity gaps in the forward
  region \cite{gap,h1flow2}. 

\item 
Remaining background is rejected by requiring no 
veto from the time-of-flight counters.

\end{itemize}

The kinematic variables are determined using 
information from the scattered positron:
$ Q^2 = 4\,E_e \, E'_e\cos^2(\theta_{e}/2)$ and
$ y = 1-(E'_e/E_e)\cdot \sin^2(\theta_{e}/2)$.
The scaling variable $x$
is then derived via $x=Q^2/(ys)$, and 
the hadronic invariant mass squared is
$W^2=sy-Q^2$.

\begin{itemize}
\item 
As the 
precision of the $y$ measurement degrades with $1/y$, 
a cut $y > 0.05$ is imposed.
 Events in which the positron is
                  poorly reconstructed, or in which an energetic photon
                  has been radiated from the incoming positron, are
                  removed by demanding that the value of $y$
                  determined using the hadrons
                  $y_h=\sum (E-p_z)/2E_e$,
                  where the sum runs over all hadronic energy deposits,
                  also be greater than 0.05.
\end{itemize}

Charged particle tracks are measured in the forward tracker and the CJC. 
They are required to originate from the primary vertex and 
their polar angle must lie between
$8^\circ$ and $155^\circ$.
 The upper limit ensures that the
                  scattered positron does not enter the sample of
                  charged hadronic tracks.
Central tracks are required to have 
a radial track length of at least 10~cm to ensure
good momentum resolution and a measured point in the inner CJC.  
Forward tracks are required to be well reconstructed 
and to have an acceptable link $\chi^2$ in the overlap 
region between the forward and central trackers.  
After having selected such high quality tracks,
the efficiency for finding a genuine primary track 
with $p_T^{\rm lab}>0.15$ GeV is better than 93\% for central tracks 
and 70\% for forward tracks.

\section{Results}

The results are presented for
the hadronic centre of mass system (CMS),
i.e.\ 
the rest system of the proton and the exchanged
boson, 
where the direction of
the exchanged boson defines the positive $z^\prime$ axis.
The event sample is divided into 9 different kinematic
bins, resulting in
three slices with almost 
constant \Qsq and varying $x$ 
(see Table~\ref{tab:et}).
The same binning is used for the measurement of
transverse energy flows
\cite{h1flow3}.

\begin{table}
\begin{center} 
\begin{tabular}{|c|c|c|c|c|c|c|}
  \hline
 Kin.bin & 
\xB $/10^{-3}$ & $\Qsq / \GeVsq$ & \av{x} $/10^{-3}$ & $\av{Q^2}/\GeVsq$ &  
 $\av{W^2}/\GeVsq$ & \#events \\
\hline
         &      &       &        &   &  & \\
0 &  0.1--10 & 5--50 & 1.14  &  18.3   &  24975 & 59463  \\
\hline
  \hline
1 &  0.1--0.2 & 5--10 & 0.16  &  7.0   &  45296 & 4853 \\
2 &  0.2--0.5 & 6--10 & 0.29  &  8.8   &  31686  & 6294\\
\hline
3 &  0.2--0.5 & 10--20& 0.37  & 13.1   &  36893  & 9564\\
4 &  0.5--0.8 & 10--20& 0.64  & 14.0   &  22401  & 7129\\
5 &  0.8--1.5 & 10--20& 1.1   & 14.3   &  13498 & 7964 \\
6 &  1.5--4.0 & 10--20& 2.1   & 15.3   &   7543 & 4720 \\
\hline
7 &  0.5--1.4 & 20--50& 0.93  & 28.6   &  32390 & 8349 \\
8 &  1.4--3.0 & 20--50& 2.1   & 31.6   &  16025 & 6752 \\
9 &  3.0--10  & 20--50& 4.4   & 34.7   &   8225 & 3838 \\
  \hline 
\end{tabular}
\end{center} 
\caption[]{
{\footnotesize
The kinematic intervals and the average $x$, $Q^2$ and $W^2$ from 
the raw data. The numbers of events surviving the selections 
are given.
}
}
\label{tab:et}   
\end{table} 

The data are corrected bin-by-bin for detector effects,
including photon conversions, secondary interactions, 
geometrical acceptance and for
decay products of particles with a lifetime greater than 8 ns
(e.g. $K^0_s\rightarrow \pi^+\pi^-$ and $\Lambda \rightarrow p \pi$),
as well as for QED radiative effects.
The corrections are determined using Monte Carlo generated
events~\footnote{
        The event generator used is DJANGO~\cite{django6}
        which simulates electroweak interactions 
         using the \mbox{HERACLES}~\cite{heracles}
        algorithm
        and QCD corrections using the ARIADNE~\cite{ariadne} program.
        The
        detector response is simulated using a program based
        on the GEANT~\cite{geant} package. The Monte Carlo events are
        reconstructed in the same manner as the $e^+p$ collision data. 
        The simulated and reconstructed events 
        obtained using DJANGO describe the raw data spectra very well. 
        Differences in the correction factors obtained
        with the alternative model LEPTO, which describes
        the data less well, are reflected in the systematic errors.}
with a full simulation of the H1 detector response. 
The detector simulation has been checked by comparing in great detail
distributions of track quality properties to the data. A visual 
scan of tracks in real and in simulated events leads to the conclusion 
that the tracker efficiency is simulated to 
an accuracy of better than 2\% for central tracks and 10\%
for forward tracks.
No spurious tracks due to noise hits have been found.
No significant difference was found between the $p_T$ spectra
of positively and negatively charged tracks.

Transverse momentum spectra are measured with the central tracker
for the pseudorapidity 
interval $1.5<\eta<2.5$, where 
\mbox{$\eta = -\ln\tan (\theta^* /2)$}, and $\theta^*$ is the angle
with respect to the virtual photon direction. 
The lower $\eta$ limit is determined
by the requirement of good acceptance in all 9 kinematic bins, and the
upper limit restricts the measurement to a region away from the 
current fragmentation region. That is where significant differences 
between the different QCD evolution scenarios are expected.

\begin{figure}[htb]
   \centering
   \vspace{-1cm}
   \epsfig{file=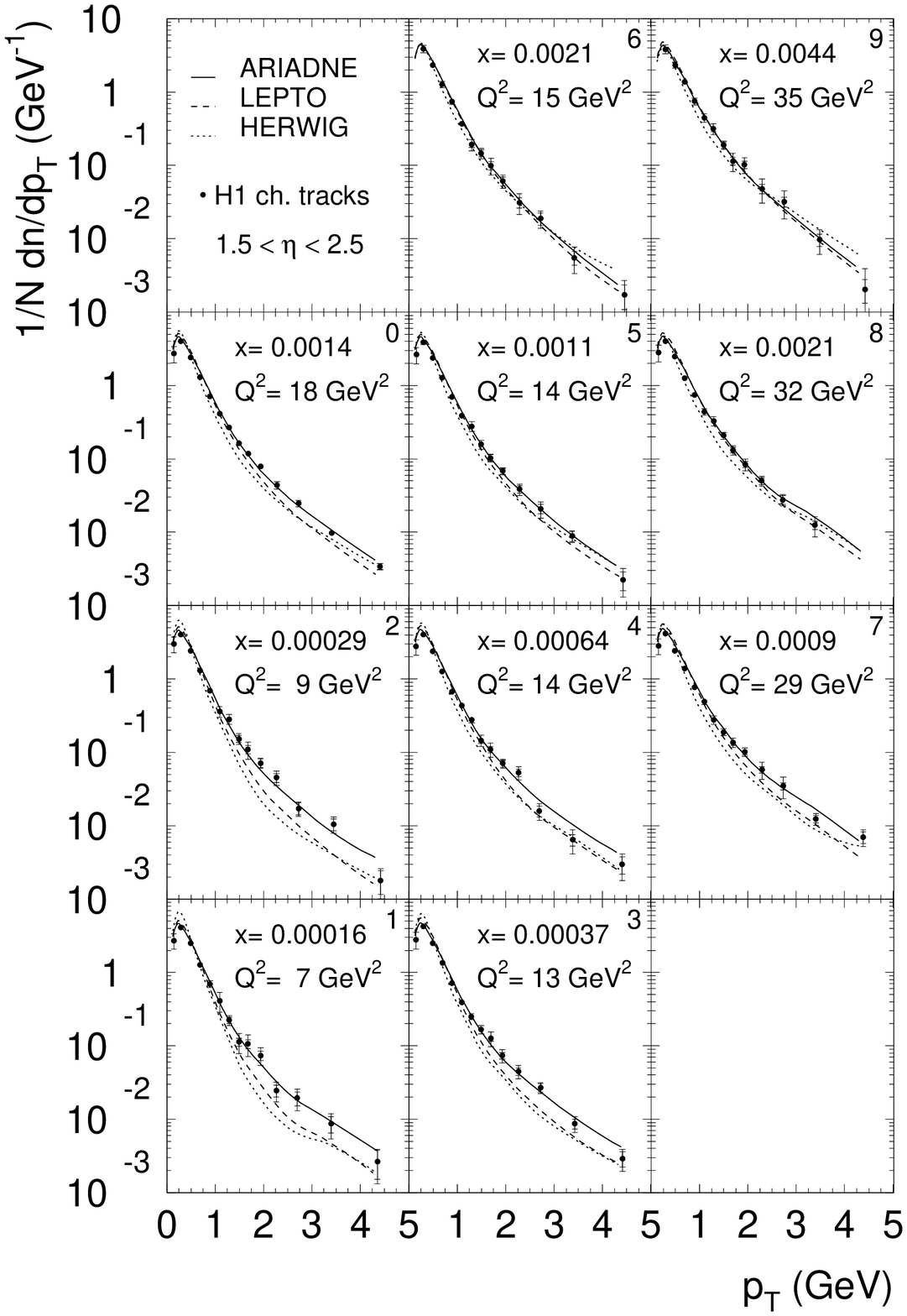,width=14cm,%
    bbllx=33pt,bblly=23pt,bburx=487pt,bbury=689pt,clip=}
  \scaption{
            The transverse momentum spectra of charged particles,
            measured in the CMS in the
            pseudorapidity interval $1.5 < \eta < 2.5$. 
            Data are shown for nine different kinematic bins
            (see Table~1) and the combined sample (bin 0).
            For comparison, the models 
            ARIADNE (full line), LEPTO (dashed) and 
            HERWIG (dotted) are overlayed. 
            The mean values of 
            \xb and \Qsq are indicated.
The inner error bars represent the statistical errors, the outer 
error bars the quadratic sum of statistical and systematic errors.
}
   \label{ptcms1}
\end{figure}

\begin{figure}[htb]
   \centering
   \vspace{-1cm}
   \epsfig{file=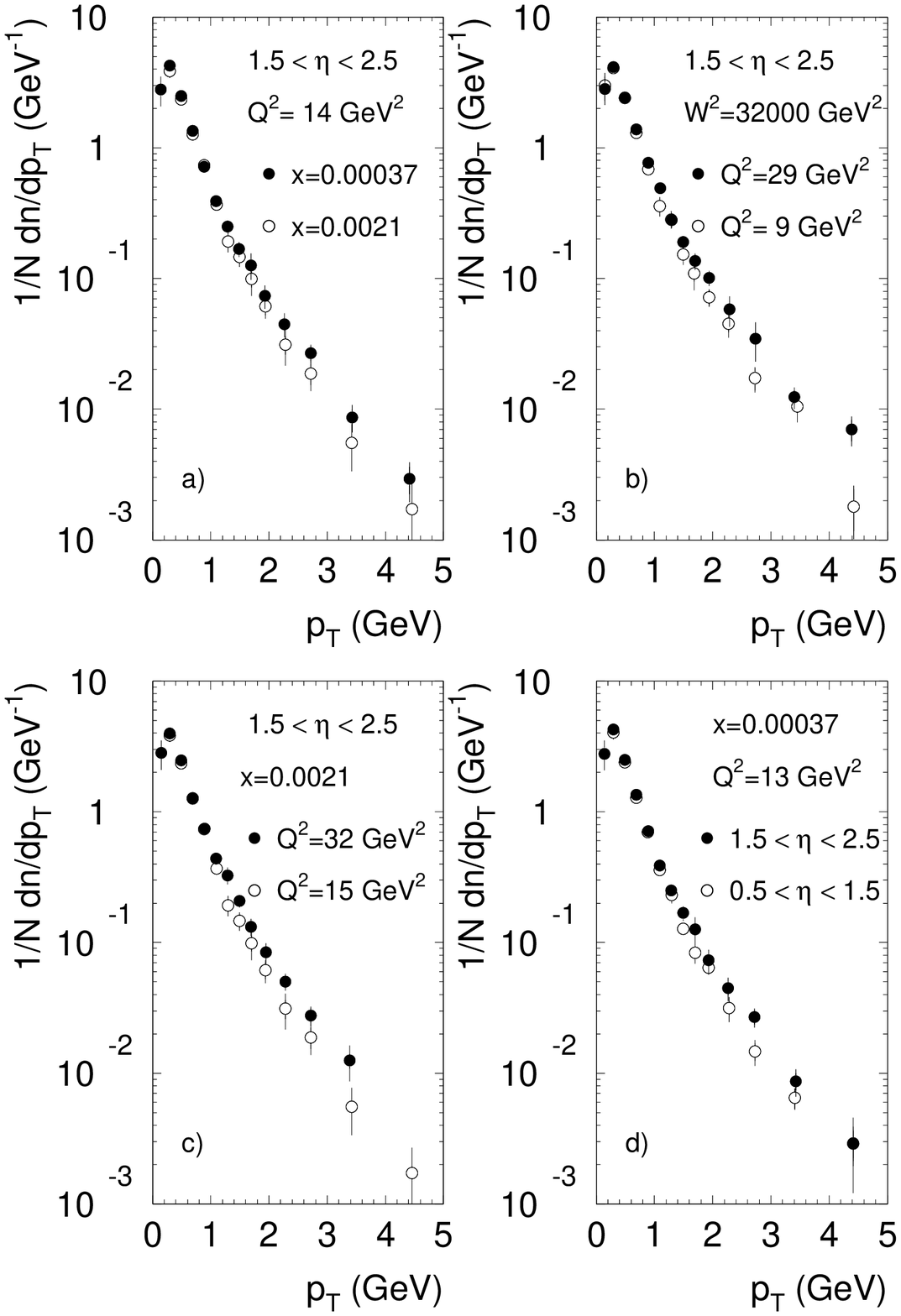,width=14cm,%
    bbllx=33pt,bblly=23pt,bburx=487pt,bbury=689pt,clip=}
   \scaption{
            {\bf a)} 
            Comparison of the \pt spectra at high \xb (bin 6)  
            and at low \xb (bin 3) for
            fixed (intermediate) \Qsq.
            {\bf b)}
            Comparison of the \pt spectra at two different \Qsq 
            values (bins 2 and 7) for fixed (large) $W$.
            {\bf c)}~Comparison
             of the \pt spectra at two different \Qsq 
            values (bins 6 and 8) for fixed (large) $x$.
            {\bf d)}
            Comparison of the \pt spectra in two different
            $\eta$ intervals in the kinematic bin 3 (low $x$,
            intermediate $Q^2$).
            The mean values of \xb and \Qsq are indicated 
            in the figures.
The error bars represent 
the quadratic sum of the statistical and systematic errors.}
   \label{ptalot}
\end{figure}

The measured spectra, 
steeply falling with $p_T$, 
are shown in Fig.~\ref{ptcms1} for the nine
kinematic bins and for the combined event sample.
All distributions shown  in this paper are 
normalized to the number of events $N$ that enter the distribution. 
The inner error bar represents the statistical error and the outer
error bar the quadratic sum of statistical and systematic errors 
(except for Fig.~\ref{ptalot}, where only the total error is shown).
We first discuss
the systematic errors
and then turn to a discussion of the results.

The correction factors applied to the original distributions
 are never larger than 2 or smaller than 0.5
and in most cases they  are between 0.9 and 1.1.
The \pt bins are much larger than the
experimental resolution and bin centre corrections are applied.
The boost to the CMS introduces a 
well understood 
change of the $p_T$ of the tracks of less than 10\% from the 
laboratory value.
The largest corrections are those necessary to compensate for event 
migration effects (contributing mainly at large $p_T$)
and track selection cuts and detector acceptance 
(contributing mainly at small $p_T$).
The main sources of systematic uncertainty are:
\begin{itemize}

\item
The model dependence of the acceptance corrections
 (ARIADNE versus LEPTO). These differ by as much as 
20\%, but in general agree to within  5\%.

\item
The track selection cuts. 
Variation of these leads to 
a 10\% systematic error
for particles with $p_T <0.5\GeV$
and for the kinematic bins at low $W,$
where measurements are close to the limit of the detector acceptance.

\item
A contamination of up to 5\% 
from photoproduction for the kinematic bins
at large $W$ \cite{gpbg}.
In order to estimate the uncertainty due to this contamination,
stricter electron quality cuts have been applied 
at the cost of statistics, 
and events with signals in the electron/photon taggers are rejected.
Differences in the result obtained with respect to 
 the standard selection are reflected in the systematic
error and are up to 30\% in kinematic bin 1, below 10\% in bin 2 and below
 5\% in the other bins.     

\item 
The uncertainty of the BEMC energy scale. This affects the boost to
the CMS. However, the effect is
negligible for all bins except for those at large $x$ and $Q^2$,
where it may give rise to an uncertainty up to 15\%.

\end{itemize}

The $p_T$ spectra shown in Fig.~\ref{ptcms1} are compared to 
the models ARIADNE, LEPTO
and HERWIG, representing the 
suppressed (LEPTO, HERWIG) and
unsuppressed (ARIADNE) gluon radiation scenarios.
At ``large'' \xb and \Qsq 
(here ``large'' means $Q^2\approx 35\,$GeV$^2$ 
and $x \approx 0.004$) 
all models provide a good
description of the data. At smaller \xb and $Q^2$, LEPTO and HERWIG
fall significantly below the data for $p_T > 1\,$GeV. ARIADNE
gives a good description of the data over the full kinematic range.
As shown in \cite{kuhlen2}, the hardness of
the \pt spectrum can be related to the 
parton radiation.
The shortfall
of the models with suppressed gluon radiation indicates that
at small \xb there is more high $k_T$ parton radiation present than is
produced by the models based upon leading log DGLAP parton showers. 
The data are well described by the ARIADNE model
 in which parton radiation is more abundant. 

In order to look for  the predicted 
hardening of the \pt spectrum with decreasing \xb
due to increasing gluonic activity,
the \pt spectra are compared in Fig.~\ref{ptalot}a 
for
the highest and lowest \xb bins (differing by a factor 6 in $x$)
of the intermediate \Qsq ($\approx 14\,$GeV$^2$) 
slice.
There is a trend consistent with the prediction of the ARIADNE model,
however not very significant.
Because $W^2 \approx Q^2/x$, 
this effect can also be understood as a $W$ dependence.

From a comparison of bins 2 and 7 (Fig.~\ref{ptalot}b), where 
$W^2$ is fixed ($\approx 32000 \rm GeV^2$) and \Qsq varies by a factor 3
(from 9 to 29 $\rm GeV^2$), it can be seen that the \pt spectrum
becomes harder with increasing $Q^2$.
When \xb is fixed ($x=0.0021$), and \Qsq increased by 
a factor 2, 
the \pt spectrum also becomes harder,
see Fig.~\ref{ptalot}c.
We have observed previously that in the current fragmentation region
the transverse energy, $E_T$, 
increases with \Qsq for fixed $W$~\cite{disgamp} and that the 
\Qsq dependence
diminishes towards the proton remnant direction.
The \et 
measured in the interval $1.5<\eta<2.5$ still increases slightly with
$Q^2$. In the central interval
$-0.5<\eta<0.5$ no \Qsq dependence of \et 
is seen for $Q^2<50\,$GeV$^2$. 
To summarize, at fixed \Qsq there appears to be 
a hardening of the \pt spectrum
with decreasing \xb and correspondingly 
increasing $W$, while at fixed \W there
is a hardening for increasing $Q^2$.

\begin{figure}[htb]
   \centering
   \vspace{-1cm}
   \epsfig{file=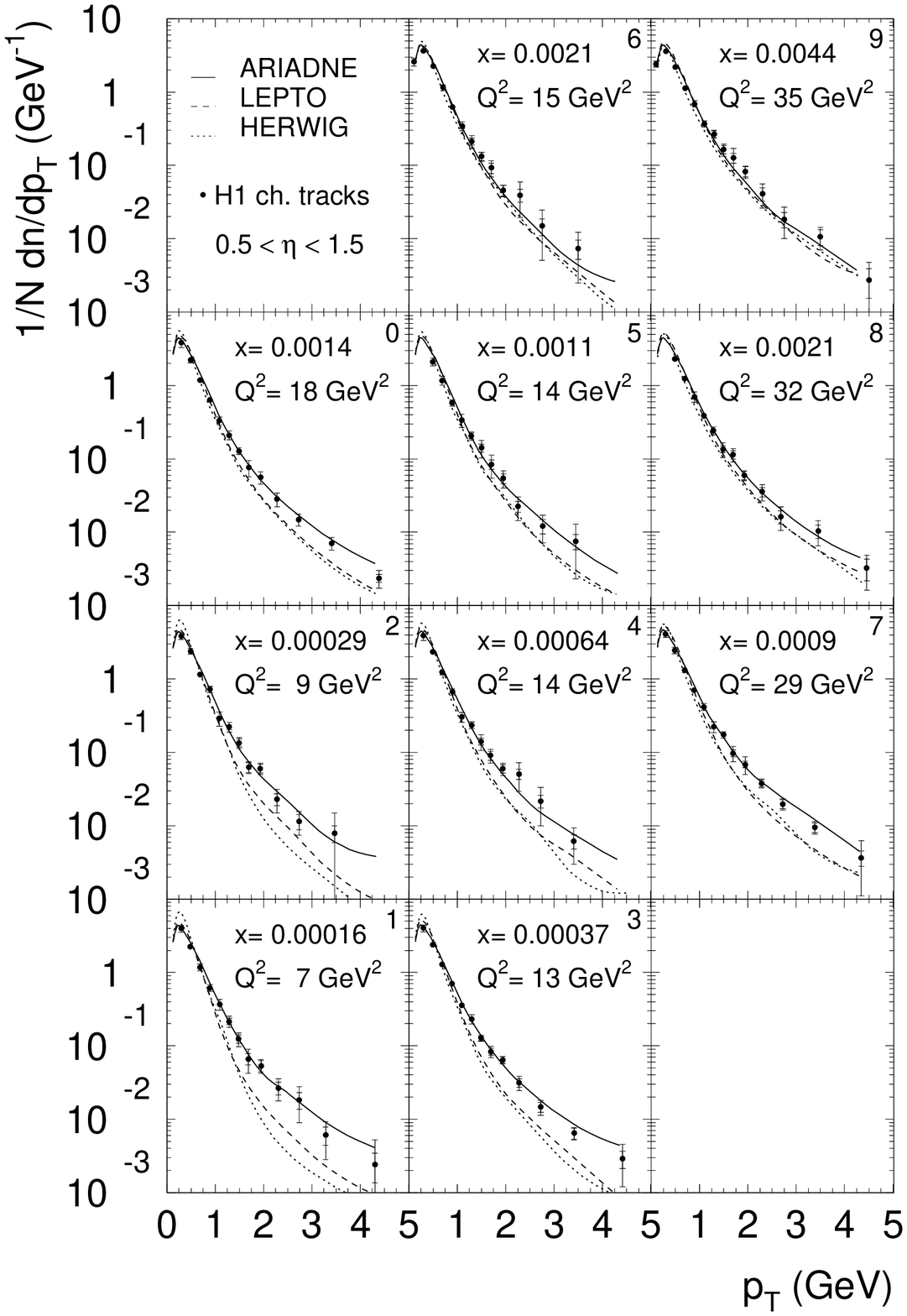,width=14cm,
    bbllx=33pt,bblly=23pt,bburx=487pt,bbury=689pt,clip=}
   \scaption{
            The transverse momentum spectra of charged particles,
            measured in the CMS in the
            pseudorapidity interval $0.5 < \eta < 1.5$. 
            Data are shown for seven different kinematic bins
            (see Table~1) plus the combined sample (bin 0).
            For comparison, the models 
            ARIADNE (full line), LEPTO (dashed) and 
            HERWIG (dotted) are overlayed. 
            The mean values of 
            \xb and \Qsq are indicated.
The inner error bars represent the statistical errors, the outer 
error bars the quadratic sum of statistical and systematic errors.}
   \label{ptcms2}
\end{figure}

Both the forward tracker and the CJC are used to measure
the \pt spectrum in the more central region 
$0.5 < \eta < 1.5$ (see Fig. 4), 
where the difference between the two types
of models becomes stronger.  Bins 6 and 9 contain roughly equal numbers
of forward and central tracks whereas the remaining bins are based on central 
tracks alone since they contain few forward tracks.  In addition to the 
sytematic errors described above, the data in bins 6 and 9 have a 
contribution from a 10\% uncertainty in the forward tracker 
reconstruction efficiency.  
The failure of the event generators
which are based on leading log DGLAP parton showers to describe  the 
hard component present in the data now becomes  more apparent.
ARIADNE describes the data well.
Comparing the \pt spectra for the two pseudorapidity intervals
in the same kinematic bin (bin 3, low $x$), it is found
that the spectrum becomes harder when one moves towards the
current direction, see Fig.~\ref{ptalot}d. 
This is not surprising, as the average \kt of the gluons emitted
in both scenarios, DGLAP and BFKL, increases along the ladder
towards the photon vertex \cite{ordering}. 
Also, radiation associated with the current system, for which the
relevant scale is 
the virtuality of the photon ($Q^2$),
is expected
to contribute more in that direction.

For the LEPTO predictions in this paper 
we modified some LEPTO parameters to improve somewhat the
description of the data over the standard setting.
The divergency cut-off parameters for the matrix element were 
changed~\footnote{
parl(8) changed from 0.01  to 0.04 and parl(9) from 1 to 4 \cite{jetset}.},
and a fragmentation parameter set tuned to the LEP data~\cite{leptune}
 was used.
The most sensitive
parameter, however,
is the virtuality cut-off below which final state radiation
is not allowed, and fragmentation takes over. 
If this parameter
would be changed from its default value of 1 GeV$^2$ to 3 GeV$^2$, 
a much better description of the \pt 
spectra for $1.5<\eta<2.5$ could be achieved.
However,
this setting would result in a poor description of the observed 
scaling violations in the 
Feynman-$x$ spectra \cite{h1flow2,zeusxf}. 
 In addition, the \pt spectra measured more   
centrally at $0.5<\eta<1.5$ are still much harder than in the
modified model. 

\begin{figure}[htb]
   \centering
   \vspace{-1cm}
   \epsfig{file=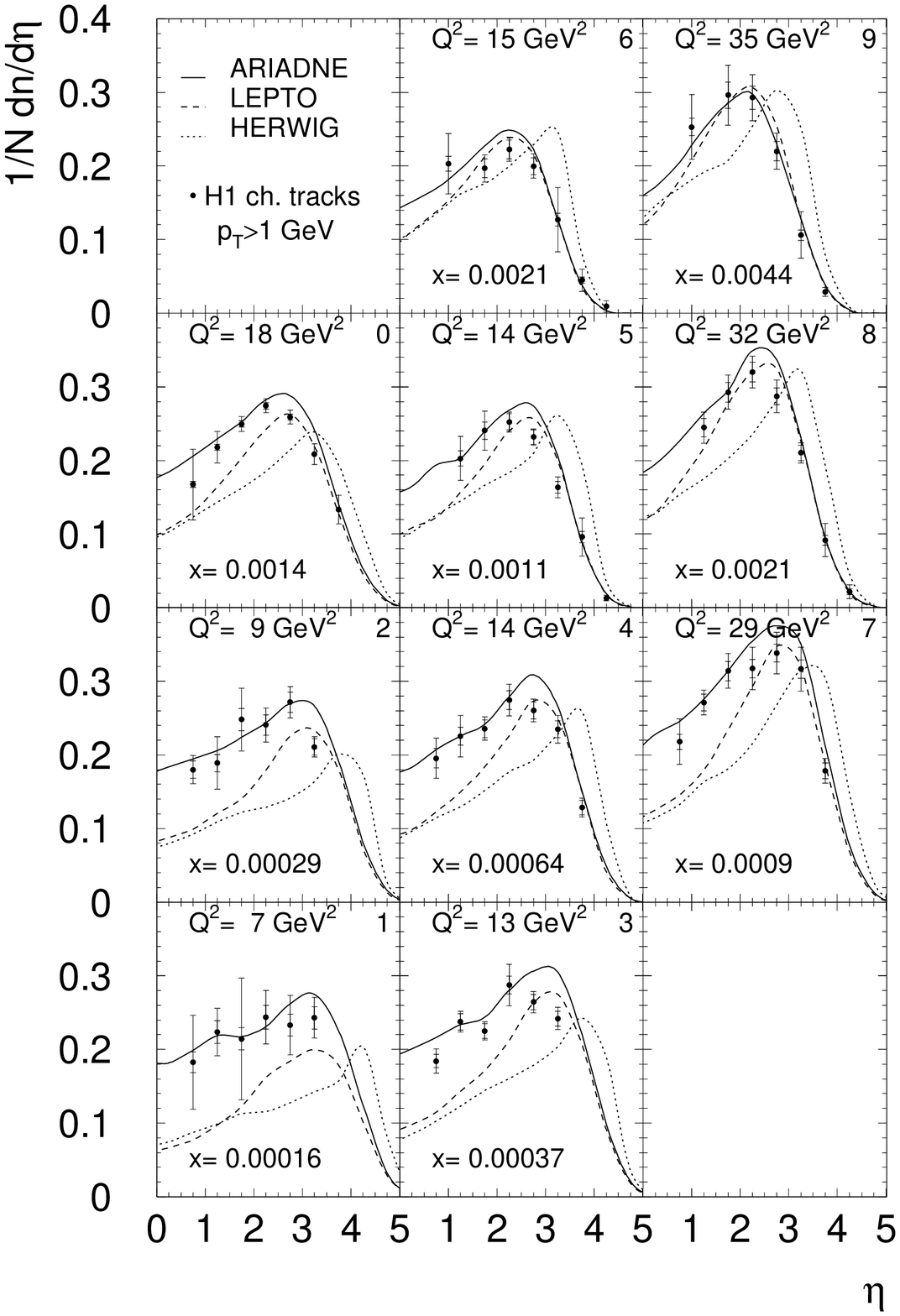,width=14cm,
    bbllx=33pt,bblly=23pt,bburx=487pt,bbury=689pt,clip=}
   \scaption{
            The CMS pseudorapidity spectra for charged particles with 
            $p_T>1$ GeV. 
            The proton remnant direction is to the left.
            Data are shown for nine different kinematic bins
            (see Table~1) plus the combined kinematic region (bin 0).
            For comparison, the models ARIADNE (full line), 
            LEPTO (dashed) and 
            HERWIG (dotted) are overlayed.
            The mean values of 
            \xb and \Qsq are indicated.
The inner error bars represent the statistical errors, the outer 
error bars the quadratic sum of statistical and systematic errors.}
   \label{dndeta1}
\end{figure}

\begin{figure}[htb]
   \centering
   \vspace{-1cm}
   \epsfig{file=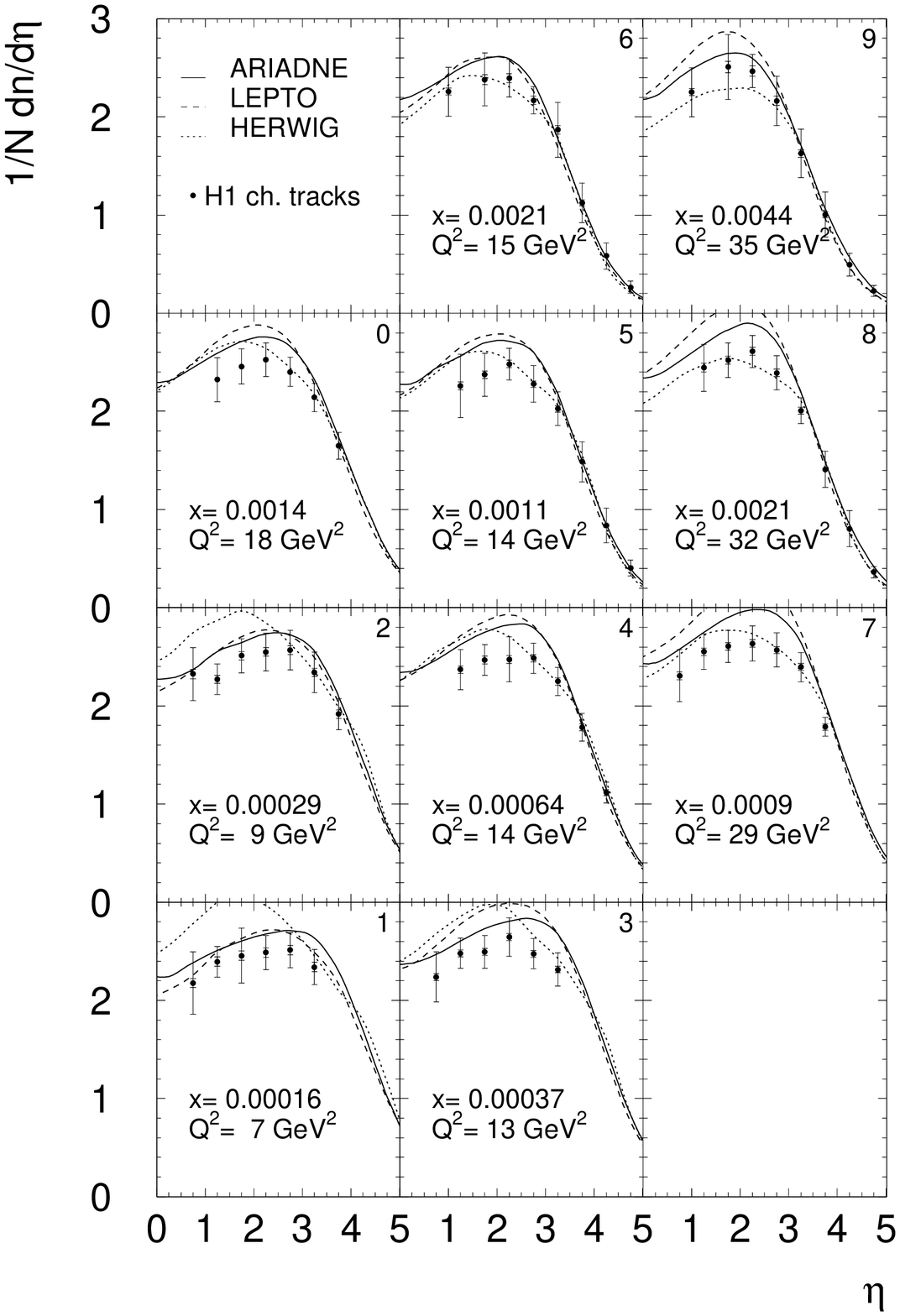,width=14cm,
    bbllx=33pt,bblly=23pt,bburx=487pt,bbury=689pt,clip=}
   \scaption{
            The CMS pseudorapidity spectra for all charged particles.
            The proton remnant direction is to the left.
            Data are shown for nine different kinematic bins
            (see Table~1) plus the combined kinematic region (bin 0).
            For comparison, the models ARIADNE (full line), 
            LEPTO (dashed) and 
            HERWIG (dotted) are overlayed.
            The mean values of 
            \xb and \Qsq are indicated.
The inner error bars represent the statistical errors, the outer 
error bars the quadratic sum of statistical and systematic errors.}
   \label{dndeta}
\end{figure}

In order to obtain a measure for the hardness of the \pt 
spectrum over the entire range of pseudorapidity accessible
to this analysis, the average multiplicity of charged particles with
$p_T >1$ GeV is measured (see Fig.~\ref{dndeta1})
as a function of pseudorapidity.
The surplus of hard particles in the data over the DGLAP-like models at
small \xb and away from the current region is obvious. 
The best description of the data is achieved by ARIADNE.
In contrast to the multiplicity of hard particles, the 
overall multiplicity, which is dominated by soft particles, 
is much better (although not perfectly) described by all models, 
see Fig.~\ref{dndeta}.
Only the HERWIG model overshoots
the data considerably at small \xb and towards the central region.

\begin{figure}[htb]
   \centering
   \vspace{-3cm}
   \epsfig{file=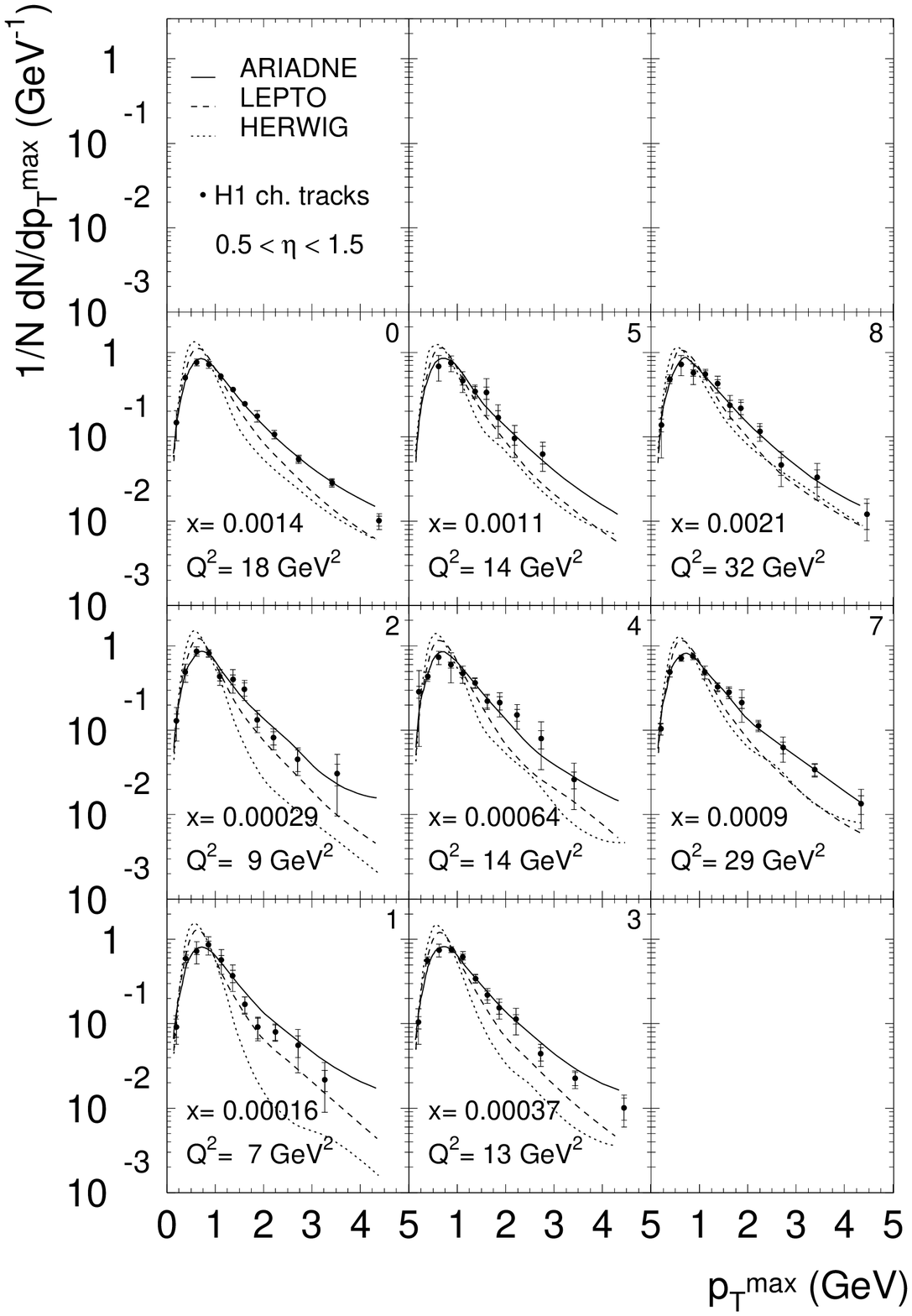,width=14cm,%
    bbllx=33pt,bblly=23pt,bburx=487pt,bbury=689pt,clip=}
   \scaption{
            The \ptmax distribution as explained in the text,
            for the interval $0.5<\eta<1.5$.
            Data are shown for seven different kinematic bins
            (see Table~1) plus the combined kinematic region (bin 0).
            For comparison, the models ARIADNE (full line), 
            LEPTO (dashed) and 
            HERWIG (dotted) are overlayed.
            The mean values of 
            \xb and \Qsq are indicated.
The inner error bars represent the statistical errors, the outer 
error bars the quadratic sum of statistical and systematic errors.}
   \label{ptmax1}
\end{figure}

The discrimination between the different scenarios
can be enhanced by selecting events with substantial 
hadronic activity 
because, in such events, there is also substantial 
partonic activity 
for the model with unsuppressed
radiation,
as opposed to the other models.
Here the transverse energy measured with the calorimeter in
the $\eta$ range from 0.0 to 2.0 is required to be larger
than 6 GeV.  
The \pt distribution of the track with the largest transverse
momentum in the range $0.5<\eta<1.5$, \ptmax, in each event is
then determined, see Fig.~\ref{ptmax1}. 
In this way one can reveal the correlation between large
\et production, which could be produced by many soft particles
in the hadronization, and high \pt particles, which
indicate hard parton radiation.
The events contain much harder particles than
are produced in the models MEPS and HERWIG.
Again, ARIADNE with enhanced parton
radiation describes the data very well.
The discrepancy between the data and ARIADNE on the one hand, 
and LEPTO and HERWIG on the other, becomes
larger for smaller \xb and \Qsq.
It is noteworthy that the colour dipole model (ARIADNE) is able to 
describe kinematic regions 
at both large and small $x$ 
simultaneously.

\section{Conclusions}

Transverse momentum spectra of charged particles in deep inelastic
scattering are measured as a function of \xb and \Qsq 
in the current and the central fragmentation regions. 
The average charged particle multiplicity is also measured
as a function of pseudorapidity.
The data are compared to QCD models incorporating different schemes for
parton evolution in the proton, 
namely both with and without transverse momentum ordering of the 
parton emissions.
The latter yields a larger amount of high $k_T$ parton 
radiation between the current and remnant systems.
 In this respect the
                   results obtained using the models
                   studied can be considered as representative of the
                   expectations for the hadronic final state
                   of the DGLAP and BFKL evolution schemes,
                   respectively.
At large $x$,  $x>0.001$, all models provide an overall 
satisfactory description of the 
measured \pt spectra.
At small \xb however and at central rapidity
the \pt distributions are significantly harder
than expected from the models with suppressed radiation.
The unsuppressed
scenario, as realized in a colour dipole radiation model,
is able to describe the data. 
The available models for the description of the hadronic
final state in deep inelastic scattering based on the conventional
leading log DGLAP equations are not able to describe the data.
The observed hard \pt spectra indicate 
hard parton radiation,
which could be provided by BFKL-type contributions.
Further interpretation of the data has to await firm QCD calculations
and a Monte Carlo model which explicitly includes the BFKL terms.

\bigskip\bigskip
\noindent
{\bf Acknowledgements.}
{\footnotesize We are grateful to the HERA machine group whose outstanding
efforts made this experiment possible. We appreciate the
immense effort of the engineers and technicians who
constructed and maintained the detector. We thank the funding
agencies for financial support. We acknowledge the support of the
DESY technical staff. We also wish to thank the DESY directorate
for the hospitality extended to the non-DESY members of the
collaboration. }
   
%
\newpage

\begin{thebibliography}{10}

\bibitem{dglap}
{ Yu. L. Dokshitzer, Sov. Phys. JETP 46 (1977) 641; \\ V.N. Gribov and L.N.
  Lipatov, Sov. J. Nucl. Phys. 15 (1972) 438 and 675; \\ G. Altarelli and G.
  Parisi, Nucl. Phys. 126 (1977) 297}.

\bibitem{bfkl}
{ E.A. Kuraev, L.N. Lipatov and V.S. Fadin, Sov. Phys. JETP 45 (1972) 199; \\
  Y.Y. Balitsky and L.N. Lipatov, Sov. J. Nucl. Phys. 28 (1978) 282}.

\bibitem{ordering}
{ J. Bartels, H. Lotter, Phys. Lett. B309 (1993) 400; \\ A. Mueller, Columbia
  preprint CU-TP-658 (1994);\\ J. Bartels, H. Lotter and M. Vogt, Phys. Lett.
  B373 (1996) 215}.

\bibitem{h1f2}
{ H1 Collab., T. Ahmed et al., Nucl. Phys. B439 (1995) 471}.

\bibitem{zeusf2}
{ ZEUS Collab., M. Derrick et al., Z. Phys. C65 (1995) 379; Z. Phys. C69 (1996)
  607}.

\bibitem{akms}
{ A.J. Askew, J. Kwieci\'{n}ski, A.D. Martin, P.J. Sutton, Phys. Lett. B325
  (1994) 21}.

\bibitem{grv}
{ M. Gl\"uck, E. Reya, A. Vogt, Z. Phys. C67 (1995) 433}.

\bibitem{h1flow2}
{ H1 Collab., I. Abt et al., Z. Phys. C63 (1994) 377}.

\bibitem{durham}
{ J. Kwieci\'{n}ski, A.D. Martin, P.J. Sutton and K.Golec-Biernat, Phys. Rev.
  D50 (1994) 217;\\ K.Golec-Biernat, J. Kwieci\'{n}ski, A.D. Martin and P.J.
  Sutton, Phys. Lett. B335 (1994) 220}.

\bibitem{h1flow3}
{ H1 Collab., S. Aid et al., Phys. Lett. B356 (1995) 118}.

\bibitem{kuhlen1}
{ M. Kuhlen, MPI-PhE/95-19 (1995), hep-ex/9508014; Proc. of the Workshop on
  Deep Inelastic Scattering and QCD - DIS95, Paris, April 1995, eds. JF.
  Laporte and Y. Sirois, p. 345}.

\bibitem{sci}
{ A. Edin, G. Ingelman and J. Rathsman, Phys.Lett. B366 (1996) 371; \\ A. Edin,
  G. Ingelman and J. Rathsman, DESY 96-060}.

\bibitem{kuhlen2}
{ M. Kuhlen, Phys. Lett. B382 (1996) 441;\\ M. Kuhlen, contribution to appear
  in the Proc. of the Workshop on ``Future Physics at HERA'', 1995-1996, eds.
  A. De Roeck, G. Ingelman and R. Klanner}.

\bibitem{lepto}
{ G. Ingelman, Proc. of the Workshop on Physics at HERA, Hamburg 1991, eds. W.
  Buchm\"uller and G. Ingelman, vol. 3, p. 1366}.

\bibitem{dipole}
{ G. Gustafson, Ulf Petterson, Nucl. Phys. B306 (1988); \\ G. Gustafson, Phys.
  Lett. B175 (1986) 453; \\ B. Andersson, G. Gustafson, L. L\"onnblad, Ulf
  Petterson, Z. Phys. C43 (1989) 625}.

\bibitem{ariadne}
{ L. L\"onnblad, Comp. Phys. Comm. 71 (1992) 15}.

\bibitem{string}
{ T. Sj\"ostrand, Comp. Phys. Comm. 39 (1986) 347; \\ T. Sj\"ostrand and M.
  Bengtsson, Comp. Phys. Comm. 43 (1987) 367; \\ T. Sj\"ostrand,
  CERN-TH-6488-92 (1992)}.

\bibitem{herwig}
{ G. Marchesini et al., Comp. Phys. Comm. 67 (1992) 465}.

\bibitem{seymour}
{ M. Seymour, Lund preprint LU-TP-94-12 (1994); \\ M. Seymour, Nucl. Phys. B436
  (1995) 443}.

\bibitem{cluster}
{ B.R. Webber, Nucl. Phys. B238 (1984) 492}.

\bibitem{bfklcdm}
{ L. L\"onnblad, Z. Phys. C65 (1995) 285; \\ A. H. Mueller, Nucl. Phys. B415
  (1994) 373;\\ L. L\"onnblad, CERN-TH/95-95}.

\bibitem{carli}
{ T. Carli, Proc. of the DIS96 Workshop, Rome 1996, ed. A. Nigro}.

\bibitem{gap}
{ ZEUS Collab., M. Derrick et al., Phys. Lett. B315 (1993) 481;\\ H1 Collab.,
  T. Ahmed et al., Nucl.Phys. B429 (1994) 477}.

\bibitem{mrsh}
{ A.D. Martin, W.J. Stirling and R.G. Roberts, Proc. of the Workshop on Quantum
  Field Theory and Theoretical Aspects of High Energy Physics, eds. B. Geyer
  and E.M. Ilgenfritz (1993) p. 11}.

\bibitem{h1nim}
{ H1 Collab., I. Abt et al., DESY H1-96-01, 
  accepted by Nucl. Instr. and Meth.}.

\bibitem{larc}
{ H1 Calorimeter Group, B. Andrieu et al., Nucl. Instr. and Meth. A336 (1993)
  460}.

\bibitem{h1pi}
{ H1 Calorimeter Group, B. Andrieu et al., Nucl. Instr. and Meth. A336 (1993)
  499}.

\bibitem{h1bemc}
{ H1 BEMC Group, J.~B\'{a}n et al., Nucl. Instr. and Meth. A372 (1996) 399}.

\bibitem{ftracker}
{ H1 Forward Tracker Group, S. Burke et al., Nucl. Instr. and Meth. A373 (1996)
  227}.

\bibitem{django6}
{ K.\ Charchula, G.\ Schuler, H.\ Spiesberger, 
  Comp. Phys. Comm. 81 (1994) 381}.

\bibitem{heracles}
{ A. Kwiatkowski, H. Spiesberger and H.-J. M\"ohring, Comp. Phys. Comm. 69
  (1992) 155}.

\bibitem{geant}
{ R. Brun et al., GEANT3 User's Guide, CERN--DD/EE 84--1, Geneva (1987)}.

\bibitem{gpbg}
{ H1 Collaboration, S.\ Aid et al., Nucl. Phys. B470 (1996) 3}.

\bibitem{disgamp}
{ H1 Collab., S. Aid et al., Phys. Lett. B358 (1995) 412}.

\bibitem{jetset}
{ T. Sj\"ostrand, Comp. Phys. Comm. 39 (1986) 347; \\ T. Sj\"ostrand and M.
  Bengtsson, Comp. Phys. Comm. 43 (1987) 367, and for JETSET~7.3, T.
  Sj\"ostrand, CERN-TH-6488-92 (1992)}.

\bibitem{leptune}
{ I.G. Knowles et al., hep-ph/9601212}.

\bibitem{zeusxf}
{ ZEUS Collab., M. Derrick et al., Z. Phys. C70 (1996) 1}.

\end{thebibliography}

%
\end{document}